# Driving towards net-zero: The impact of electric vehicle flexibility participation on a future Norwegian electricity system

Tobias Verheugen Hvidsten[a,*], Maximilian Roithner[a], Fred Espen Benth[b], Marianne Zeyringer[a]

[a] Department of Technology Systems, University of Oslo, Norway.

[b] Department of Mathematics, University of Oslo, Norway.

*Corresponding author. E-mail: t.v.hvidsten@its.uio.no

## Abstract

Electric vehicle batteries have a proven flexibility potential which could serve as an alternative to conventional electricity storage solutions. EV batteries could support the balancing of supply and demand and the integration of variable renewable energy into the electricity system. The flexibility potential from electric vehicles, in distinction to conventional battery storage, depends on the vehicle user's willingness and opportunity to make their vehicle available for flexibility. This rate of participation is often not considered in studies, despite the impact electric vehicle flexibility could have on the electricity system. This work presents a modelling study of the Norwegian electricity system, demonstrating how a future net-zero electricity system can benefit from electric vehicles in terms of integrating renewables and balancing supply and demand, while considering the rate of participation. Our findings show electric vehicles' potential to eliminate the need for stationary battery storage with just 50% participation in vehicle-to-grid. We find that the flexibility of electric vehicles contributes to relative reductions in the total cost of the electricity system by almost 4% and 15% assuming 100% participation in flexible charging and vehicle-to-grid, respectively.

## 1  Introduction

The Norwegian electricity supply is already highly renewable with 95% of the electricity used in 2023 coming from renewable sources (NVE, 2024a). This is mainly due to the vast hydropower resources, alone covering 83% of electricity consumption in 2023. Norway's challenge towards 2050 is thus not to substitute current fossil fuel-based electricity generation as in many other countries, but to meet the increasing electricity demand with renewable energy. For Norway to reach its aim of reducing greenhouse gas emissions by 90 to 95% compared to 1990 (Lovdata, 2017) several sectors, mainly transport and industry (Miljødirektoratet, 2024), need to substantially increase electrification by 2050. Statnett, the Norwegian transmission system operator (TSO), predicts electricity demand to rise from 140 TWh in 2022 to between 190 TWh and 300 TWh by 2050 (Statnett, 2023). In 2021 Norway had a total electricity production of 157 TWh, the highest in the statistics going back to 1950 (Statistics Norway, 2024b). This indicates that additional electricity generation is necessary in all of the TSOs demand scenarios if Norway is to keep producing more electricity than is consumed, on a yearly basis. Despite large technical potential for additional hydropower in Norway, nature protection limits it to 23 TWh (M. E. Henriksen et al., 2020). New generation capacity to meet the growing demand is therefore expected to come mainly from variable sources of renewable energy, such as wind and solar (photovoltaic (PV)) power (Bjørndalen et al., 2023). This introduces more variability in the system, as their generation is weather dependent. Additional sources of flexibility are thus important assets in the electricity system to integrate the increasing generation from variable renewables. Flexibility can be provided in several ways, the most common is dispatchable generation, for which hydro power with reservoirs is an excellent example. Other options include electricity storage such as batteries or hydrogen, and demand side measures such as curtailing, increasing and shifting loads (Morales-España et al., 2022).

One of the sectors which is projected to drive up the electricity demand in Norway towards 2050 is transport, with several targets for electrification, including only selling zero-emission passenger cars by 2025 (Norwegian Ministry of Transport, 2021). The Norwegian TSO has projected that by 2050 the transport sector could require about 26 TWh of electricity (Statnett, 2023), up from 3 TWh in 2022. When electrifying the vehicle fleet, carbon emissions are mitigated, but at the same time electricity demand increases. This increase must be met by generation from renewables to prevent moving emissions from one sector to another. In the coupling of the transport sector and the electricity system, electric vehicle (EV) charging could be seen as just an additional load. Delmonte et al. (2020) find that most EV users prefer to fully charge their EV when they come home in the early evening. However, such uncontrolled charging behaviour could aggravate peak demand, as discussed by several studies (Crozier et al., 2020; Heuberger et al., 2020; Mangipinto et al., 2022). This again could result in a need for increased generation capacity. However, when EVs are not in use, which could be as much as 95% of the time (Gong et al., 2024), they could provide flexibility to the system. Firstly, EV charging can be shifted temporally to times more beneficial to the electricity system. Moreover, with the battery storage capacity expected to be employed in the EV fleet there is an opportunity to utilise EV batteries as an electricity storage solution. This is enabled by bidirectional charging, also known by the term vehicle-to-grid (V2G) which we will use throughout this paper. With this EVs can be charged in hours of excess electricity generation, and electricity not needed for driving can be released to cover other loads at times of high demand. EVs could consequently contribute to the integration of variable renewable energy. This flexibility from EVs depends both on the user's willingness and opportunity to participate. Certain user groups might be less likely to actively participate and can be referred to as hard-to-reach energy users (Rotmann et al., 2020).

This work considers the coupling of the electricity system and the transport sector as the latter is electrified. The potential role of EVs as a source of flexibility and electricity storage in a future Norwegian electricity system is investigated. Through exploring two forms of electric vehicle flexibility, namely flexible charging and V2G, we investigate how EV user's willingness and opportunity to participate impact the potential flexibility and storage from EV batteries. Through this we explore the degree to which hard-to-reach EV user participation impacts the system. This work addresses the following research question: How is the design and cost of a future Norwegian net-zero electricity system affected by the share of electric vehicle users participating in flexible charging or vehicle-to-grid?

Energy and electricity system models are widely used to design future least cost systems that meet climate targets, generating insights that inform energy and climate policy (DeCarolis et al., 2017). There are however few studies that use an electricity system model to study the impact of flexible charging of EVs and V2G on future electricity system design and operation: In a study for Switzerland, Syla et al. (2024) investigate the impact of flexible charging under different EV adoption rates on the cost-optimal design of the electricity system. They show that flexible charging mitigates some of the battery storage needs induced by EV uptake and supports more PV capacity in the system. This study, however, does not include V2G as a flexibility option. In another study, Guéret et al. (2024), using the DIETER electricity system model, optimise the German electricity system in 2030 considering the flexibility from EVs. Though they focus on the transition to carsharing, they also show both smart and bidirectional (V2G) charging of EVs benefits the system in terms of lower total system costs. The study however focuses on a shorter-term perspective only until 2030, which means a lower share of EVs in the system. Xu et al. (2023) develop a simulation model and find that the worldwide need for short-term electricity storage could be met with V2G in several scenarios by 2050. They also consider the participation in V2G and find participation rates as low as 12% to be sufficient to meet the demand depending on the scenario and storage needs.

No identified study optimises the design of the future Norwegian electricity system while including flexibility from EVs. Nagel et al. (2024) apply the energy system model Balmorel to optimise the operation, but with generation capacities defined exogenously, of the 2040 Norwegian electricity system. They find V2G to reduce the operational costs of the system, help integrate renewable generation and decrease curtailment. High potential for demand side flexibility is identified in Norway, with electrified heating identified as the main source (Kirkerud et al., 2021; Söder et al., 2018). In a review paper, Söder et al. (2018) argue that due to the flexible hydropower resources in Norway the need for demand side flexibility will be limited. These findings are also supported by Kirkerud et al. (2021), who despite this find, using the Balmorel energy system optimisation model, that demand side flexibility decrease the battery storage demand in Norway. Another study by Ahang et al. (2023) applies the energy system optimisation model TIMES and finds that demand side flexibility increases the value of the flexibility from hydropower, through exports and imports. Neither Ahang et al. (2023) or Kirkerud et al. (2021) include flexibility from EVs in their assessments. Their findings nonetheless motivate the assessment of the future role of flexibility from EVs in Norway.

EVs will only be able to provide flexibility to the electricity system if the user is willing and able to do so. The literature shows that there are numerous constraints to EV users' willingness and opportunity to participate. EV users facing such constraints are difficult to engage in participating actively and can be identified as hard-to-reach energy users (Rotmann et al., 2020). Three groups of hard-to-reach energy users in the residential sector are commonly mentioned in the literature: Vulnerable households, high-income households and renters (Ashby et al., 2020; Rotmann et al., 2020). Vulnerable households (e.g. lower income (Ashby et al., 2020) or elderly (Szulecki et al., 2024)) often lack the financial means and knowledge to actively participate in the energy system (Ashby et al., 2020; Standal et al., 2023). For instance, participation in EV flexibility is dependent on having the financial resources to afford the high upfront investment costs of EVs and charging infrastructure (Libertson, 2022). Sørensen et al. (2021) find the potential for flexibility to be greater for EVs with access to private charging points compared to EVs charged at shared chargers, due to the length of time the EVs can be connected to the charger. They also point out that having access to charge points with higher (dis)charging power allow for more flexibility, as electricity can be moved more quickly between the vehicle and the grid. This illustrates the infrastructure required to enable flexibility. To reach full potential, all EVs would need access to high power private charge points. High-income households can lack interest in participation as economic incentives are less impactful (Ashby et al., 2020). While renters, as they don't own the property, might experience uncertainty around who should pay for the investments to enable participation in flexibility (Ashby et al., 2020; Standal et al., 2023). Various other barriers to participation in EV flexibility are also found in the literature. For instance, Mehdizadeh et al. (2024) find concerns around the capacity and degradation of EV batteries as common barriers to V2G. Delmonte et al. (2020) find that the most common concern around flexible charging is the risk of the EV not being sufficiently charged in case of it being needed due to an unexpected event or emergency. Nevertheless, there are also many motivating factors for participation. For instance, I. M. Henriksen et al. (2021) present a case study on the motivations for smart charging of EVs in Norway. They identify several motivations including faster and safer charging with EV home chargers, charging when electricity prices are low, and preheating of the battery and the vehicle to preserve battery health and enhance comfort. According to Bailey and Axsen (2015) younger people are more positive towards flexibility options and are more likely to participate, indicating promise to the future uptake of these options and their role in the transition to a net-zero electricity system towards 2050. Despite the vast literature on barriers to participation in the electricity system, Sovacool et al. (2018) find that modelling studies usually assume that all EV users participate in V2G, identifying an important research gap.

Based on the literature we identify three gaps: First, an analysis of impacts of V2G for a net-zero electricity system is lacking. To the to the best of our knowledge Guéret et al. (2024) present the only study analysing the potential effects of V2G on future electricity system design and cost. However, their study focuses on a shorter time perspective until 2030. Second, there is a gap in the understanding of how EV flexibility would impact the optimal design of the Norwegian electricity system. Norway is an interesting case study because of the already high uptake of EVs, which makes their potential for providing flexibility significant. Flexible generation from hydropower is abundant in Norway and currently reduces the need for other sources of flexibility. However, increasing demand in the future will require more variable generation to be installed and in turn increase the need for flexibility. Third, no study incorporates the rate of participation in EV flexibility in an electricity system optimisation model. It is important to take the rate of participation into account as the prevalence of hard-to-reach energy users creates uncertainty about the role which EV flexibility options could play in future electricity systems.

To close the first and second gap, we illustrate how the storage capacity of EVs can provide flexibility, both via flexible charging and V2G, to integrate variable renewables in a fully decarbonised Norwegian electricity system in 2050. We therefore develop an EV module as an add-on to the open source high spatial and temporal resolution electricity system model (highRES) (Price & Zeyringer, 2022). This addition allows the model to simultaneously optimise investments in the electricity system and the operation (balancing supply and demand) of the EVs in conjunction with the rest of the system. This means that the EV flexibility is optimised as part of the electricity system, within the constraint that EV users' driving habits are not impacted. We apply this updated model to investigate not only how EV flexibility impacts the operation of the Norwegian system, as covered by the literature, but also how the optimised electricity system design changes with the additional flexibility provided by EVs when enabling flexible charging and V2G. To close the third gap, we investigate the effect hard-to-reach consumers' willingness and opportunity to participate in flexible charging and V2G might have on the future electricity system design. To do so we optimise a future Norwegian electricity system under different degrees of consumer participation in both flexible charging of EVs and V2G.

The paper is structured as follows. Section 2 outlines the methodology of the paper. It starts with a general description of the electricity system model and the general assumptions in that model in Subsection 2.1. The development and implementation of the electric vehicle module, which allows for the modelling of flexible charging and V2G, is covered in Subsection 2.2. Finally, Subsection 2.3 provides an overview of the scenarios modelled. The results and discussion, covered by Section 3, start with an analysis of the impacts of EV flexibility on the balancing of supply and demand in Subsection 3.1. The total system cost of the modelled system for the various scenarios is discussed in Subsection 3.2, and Subsection 3.3 details the optimised electrify system designs for the different scenarios. We end the paper by providing our conclusions as well as reviewing the limitations of our work and suggesting potential areas of further research in Section 4.

## 2 Methodology

### 2.1 The electricity system model

To model a future Norwegian electricity system in the year of 2050 we employ the high spatial and temporal resolution electricity system model (highRES) (Price & Zeyringer, 2022). The objective of the model is to minimise the total annualised cost of the system, while balancing supply and demand at an hourly temporal resolution. To be able to investigate the role of EVs in a future Norwegian electricity system the model is extended with a module which models the flexibility of EVs and their interaction with the rest of the electricity system. Details on this implementation is found in section 2.2. Following are some general details on the model and the assumptions in it:

The target year for optimisation of the Norwegian electricity system is chosen to be 2050. This year is chosen as this is a common year for climate targets, including Norway's. Following Norway's emission targets (Lovdata, 2017), a slightly more ambitious zero-emission target is assumed for the electricity system. This also means that investments in the system are limited to technologies without direct carbon emissions. These are renewable electricity generation from solar, and onshore and offshore wind. For offshore wind we distinguish between two types, bottom mounted, which we throughout this paper refer to as fixed, and floating offshore wind. New hydropower is not modelled due to the constraints discussed in the introduction. Additionally, the model can invest in stationary battery storage and transmission capacity between neighbouring model zones. The model zones correspond to the Norwegian counties pre-2024, see Figure 5. Transmission capacity between model zones is set to be at minimum equal to the transmission capacities of the current Norwegian electricity system. In addition, the model has the option to freely invest in increasing the capacities of the existing transmission lines. Following the target year, estimated technology costs for 2050 are applied for onshore wind, fixed offshore wind, floating offshore wind, solar PV (Danish Energy Agency, 2024) and stationary battery storage (Danish Energy Agency, 2023). The costs from the Danish Energy Agency include an offshore turbine category which we assume to represent fixed offshore wind. Wiser et al. (2021) find the change in levelized cost of electricity in 2050 compared to the current (2019) level for fixed offshore wind to be 49% and 40% lower for fixed and floating offshore wind respectively. Based on these results the cost of floating offshore wind relative to fixed offshore wind in 2050 would be 17.6% higher. To obtain costs for floating offshore wind we apply these findings, assuming the same difference for all cost components (capital expenditure, variable operation and maintenance, and fixed operation and maintenance).

We model only the Norwegian electricity system. Without also modelling all regions connected to Norway interconnections cannot be accurately represented. Due to this limitation imports and exports are constrained to zero. With this approach the modelled system cannot profit from exporting excess generation from variable renewables, which in this case is only curtailed. Nonetheless, initial runs of the model, with a simplified approach including imports at a fixed price and interconnection capacities set according to current net transfer capacities between Norway and neighbouring countries (ENTSO-E, 2024) showed that model results are sensitive to imports. With imports the model only makes minor investments into stationary battery storage (less than 2 GWh). Compared to this, constraining imports to zero the model invests significantly in stationary battery storage (about 71 GWh). We see that imports can mostly replace batteries' role in the balancing supply and demand. The total level of imported electricity is comparable to import levels in Norway in recent years (Statistics Norway, 2024b). Only 7.34 TWh of imports are seen, which is about 2.8% of the applied 2050 demand. However, with this implementation of imports at a fixed price, imports can rather be viewed as a dispatchable generator or energy storage with a fixed cost for the electricity supplied. The strategic behaviour in the electricity market is not reflected and it can rather be seen as a proxy where electricity can be bought at a high cost when demand cannot be met. On the other hand, Price et al. (2023) find transmission expansion to be a cost-effective solution to integrate variable renewable energy. It reduces the need for energy storage, but as noted by the authors, depending on electricity imports might not be publicly accepted, providing an argument for designing electricity systems to not be dependent on imports.

The weather and demand year used as input to the model is 2010, which was chosen as it was a difficult weather year for the Norwegian electricity system. It was an unusually cold year and had less precipitation than normal, especially during the winter and the autumn (Iden et al., 2011; Meteorologisk institutt, 2024). This resulted in high electricity demand and low production from hydropower (Statistics Norway, 2024b), due to respectively low temperatures and less precipitation leading to reduced inflow to hydropower. The hydro reservoir statistics from NVE (2024b) also shows that the fill rate in 2010 was below the median throughout the year, and the last seven weeks of the year had the lowest fill rates recorded to date. Since hydropower is the main source of electricity generation in Norway, the low generation and high demand signify that 2010 was a difficult weather year for the system. Electricity demand, apart from that modelled through the EV module described in section 2.2, is based on electricity demand time series from Frysztacki et al. (2022). We apply the demand time series for the year 2010, the same year as the weather data. This demand is taken as the baseline on top of which changes in electricity demand towards 2050 is modelled. Electricity demand in Norway in 2050 is based on the projections of the Norwegian TSO (Statnett, 2023). They project electricity demand to increase from 140 TWh in 2022 to 260 TWh, including all electric transport demand, in their high consumption prognosis for 2050. A description of the distribution of the increase in demand between the system's different sectors, as well as how the growing demand of each sector is spatially and temporally distributed, is given in Appendix A.

## 2.2 Electric vehicles: model implementation and data

Electric vehicles are in general everything from electric bicycles to electric airplanes. This work focuses on the subgroup of electric vehicles which is battery electric passenger cars. When using the term electric vehicles in the reminder of this paper we refer to this specific subgroup rather than the broader meaning of the term. The electrification of the transport sector interlinks it with the electricity system. This work includes the general electricity demand from electrified transport (see Appendix A), but for electric passenger cars we take one step further by representing them not only as additional demand but also as possible source of flexibility and energy storage. We model two different forms of flexibility from electric vehicles: Flexible charging and V2G. While there are several charging strategies for electric vehicles which can be applied to utilise the energy storage capabilities of EV batteries, as detailed by Gong et al. (2024), in this study we focus on V2G. V2G can support the balancing of the system by charging and discharging when it is optimal for the system. In general, what would be optimal is to charge when prices are low and discharge when prices are high.

*Table 1: List of nomenclature relating to the EV implementation.*

| **Scalars** | |
|---|---|
| $f_{EV}$ | Fraction of vehicles participating in flexibility. |
| $e_{EV,capacity}$ | Battery storage capacity per EV [MWh]. |
| $p_{EV,discharge}$ | Power with which each EV can discharge [MW]. |
| $p_{EV,charge}$ | Power with which each EV can charge [MW]. |
| $\eta_{charge}$ | Efficiency when charging EV batteries. |
| $\eta_{discharge}$ | Efficiency when discharging EV batteries. |
| $SOC_{min}$ | Minimum state of charge. |
| $SOC_{max}$ | Maximum state of charge. |
| **Parameters** | |
| $h$ | Model timesteps [hours]. |
| $z$ | Model zones. |
| $n_{EV}(z)$ | Number of vehicles per zone. |
| $e_{EV,static}(h)$ | EV charging demand per vehicle for uncontrolled charging [MWh]. |
| $e_{EV,driving}(h,z)$ | Electricity used while driving per car [MWh]. |
| $c_{EV}(h)$ | Fraction of cars connected to the grid. |
| **Variables** | |
| $E_{EV}(h,z)$ | Energy stored in electric vehicle batteries [MWh]. |
| $E_{EV,discharge}(h,z)$ | Energy discharged from electric vehicle batteries [MWh]. |
| $E_{EV,charge}(h,z)$ | Energy charged to electric vehicle batteries [MWh]. |

The formulation of the EV flexibility is adapted from Morales-España et al. (2022). The main addition to the formulation of Morales-España et al. is the inclusion of a parameter representing the fraction of electric vehicles participating in flexibility, $f_{EV}$. The energy stored in electric vehicle batteries, $E_{EV}(h,z)$, at each hour, $h$, and in each model zone, $z$, is given by

$$E_{EV}(h,z) = E_{EV}(h-1,z) + E_{EV,charge}(h,z) \cdot \eta_{charge} - \frac{E_{EV,discharge}(h,z)}{\eta_{discharge}} - E_{EV,driving}(h,z), \quad (1)$$

where $E_{EV}(h-1,z)$ is the energy stored in the previous time step. $E_{EV,charge}(h,z)$ and $E_{EV,discharge}(h,z)$ represent the energy charged and discharged, respectively, to the EV batteries in each hour and zone. $\eta_{charge}$ and $\eta_{discharge}$ are the charging and discharging efficiency of the EV batteries. $E_{EV,driving}(h,z)$ is the battery discharge when driving, which depends on the number of electric vehicles

$$E_{EV,driving}(h,z) = e_{EV,driving}(h,z) \cdot n_{EV}(z) \cdot f_{EV}, \quad (2)$$

where $e_{EV,driving}(h,z)$ is the electricity discharged while driving on average for an individual car. $n_{EV}$ is the number of EVs in each zone of the model. The total amount of energy stored in all EV batteries is limited by the aggregated battery capacity of all EVs

$$E_{EV}(h,z) \leq e_{EV,capacity} \cdot n_{EV}(z) \cdot f_{EV}, \quad (3)$$

where $e_{EV,capacity}$ is the battery storage capacity of each individual EV. The hourly charging and discharging of the EV batteries is limited by the power capacity of the batteries

$$E_{EV,charge}(h,z) = n_{EV}(z) \cdot f_{EV} \cdot p_{EV,charge} \cdot c_{EV}(h), \qquad (4)$$

$$E_{EV,discharge}(h,z) = n_{EV}(z) \cdot f_{EV} \cdot p_{EV,discharge} \cdot c_{EV}(h), \qquad (5)$$

where $c_{EV}(h)$ is the fraction of the EVs connected to the grid at any hour. $p_{EV,charge}$ and $p_{EV,discharge}$ is the power with which each EV can charge and discharge respectively. Finally, to preserve the battery lifetime, the state of charge (SOC) of the batteries is constrained

$$E_{EV}(h,z) \geq SOC_{min} \cdot e_{EV,capacity} \cdot n_{EV}(z) \cdot f_{EV}, \qquad (6)$$

$$E_{EV}(h,z) \leq SOC_{max} \cdot e_{EV,capacity} \cdot n_{EV}(z) \cdot f_{EV}, \qquad (7)$$

where $SOC_{min}$ and $SOC_{max}$ is the minimum and maximum allowed state of charge of each EV battery. In addition, we model the EV battery storage to be cyclical, similarly to other storage options in the model, such as stationary batteries and hydro reservoirs. This means that in the first hour, $h = 0$, the stored energy in the previous hour, $h - 1$, is the stored energy at the hour $h = 8759$ which is the last hour of the year (given that the number of timesteps equals the number of hours in a year, which is 8760). In practice what this means is that the model operates the storage under the constraint that the storage level at the start of the year must follow from the storage level at the end of the same year. In a real system the storage level at the start of the year is linked to the storage level at the end of the previous year. We cannot replicate these real-world conditions as we are only modelling a single year. However, we apply cyclical storage with the aim to improve this compared to setting an arbitrary storage level at the start of the year.

To link this EV module to the electricity system model the supply-demand balance equation is modified to include the demand from EVs in addition to the supply when allowing for V2G. In simple terms the supply-demand balance equation establishes that for each model zone and hour the supply, from generation, storage, and transmission, must equal demand, including transmission to other zones and charging of storage. In the initial supply-demand balance equation

$$E_{demand}(h,z) = E_{gen}(h,z) + E_{trans,in}(h,z) - E_{trans,out}(h,z) + E_{store,gen}(h,z) - E_{store,charge}(h,z), \qquad (8)$$

we have for every hour and zone the electricity demand, $E_{demand}(h,z)$, the generation from all technologies, $E_{gen}(h,z)$, the electricity transmitted to and from neighbouring zones, $E_{trans,out}(h,z)$ and $E_{trans,in}(h,z)$, and the electricity charged to and supplied from storage, $E_{store,charge}(h,z)$ and $E_{store,gen}(h,z)$. From EVs we get the following terms added to the right-hand side of equation 8

$$+ E_{EV,discharge}(h,z) - E_{EV,charge}(h,z) - \frac{\left(e_{EV,static}(h) \cdot n_{EV}(z) \cdot (1 - f_{EV})\right)}{\eta_{charge}}, \qquad (9)$$

where the two first terms give the supply and demand respectively from EVs participating in flexibility, while the final term gives the electricity demand from the EVs not participating. In the case of no participation in EV flexibility, $f_{EV} = 0$, the two first terms will be zero and the third term will encompass all demand from EVs. In the opposite case with all EVs participating in flexibility, $f_{EV} = 1$, the final term will be equal to zero.

The implementation of EV flexibility in the model as described above is general and could be applied in other electricity system models, covering other countries and areas. While the model implementation is applicable to other areas, the EV input data applied, which will be detailed in the remainder of this section, is specific to Norway. It would have to be adjusted to accurately represent the driving behaviour elsewhere. The fraction of EVs participating in flexibility, $f_{EV}$, can take any value between 0% and 100%, which is set depending on the scenario (see the description of scenarios in Section 2.3). When a fraction of the EV fleet shift charging to times of the day that are beneficial for the system or even send electricity back to grid, the remaining fleet is modelled with uncontrolled charging, that is charging without flexibility. For the fraction of the total EV fleet which at any hour throughout the year is connected to the grid, $c_{EV}(h)$, we apply data from Sørensen et al. (2021) who conducted a case study on EV charging and reported the hourly share of EVs connecting and disconnecting to private chargers on weekdays and weekends. The hourly change in the fraction of EVs connected, $\Delta c_{EV}(h)$, is calculated as the difference between the fraction of cars disconnecting and connecting. The fraction of EVs connected to the grid at any hour is then

$$c_{EV}(h) = c_{EV}(h-1) + \Delta c_{EV}(h), \qquad (10)$$

where $c_{EV}(h-1)$ is the fraction of EVs connected in the previous hour. To ensure that the fraction throughout the year stays within the range of zero and one, the initial fraction, $c_{EV}(0)$, must be between approximately 0.52 and 0.96. We here arbitrarily assume 0.75 somewhere in the middle of this range. With this, at any hour throughout the year the minimum share of cars connected is approximately 24% and the maximum is about 79%.

Participation in flexibility entails that when a vehicle is connected to a charging point it is accessible to be charged flexibly and potentially also be operated as an energy storage option. Figure 5 shows the assumed number of EVs in each model zone for 2050, $n_{EV}(z)$. The number of EVs in 2050 is based on projections from the Norwegian Centre for Transport Research (Fridstrøm, 2019) on the future Norwegian vehicle fleet. First of all, they expect the size of the car fleet to increase, reaching about 3.46 million passenger cars by 2050. Of these it is projected that the vast majority, about 3.25 million, are battery electric in a scenario in line with the goal of only selling zero emission passenger cars starting from 2025 (Norwegian Ministry of Transport, 2021). The zonal distribution of the EVs (see Table 2 in Appendix B) is based on the current number of registered private cars in each county in Norway in 2023 (Statistics Norway, 2024a). The number of EVs and how many of these are connected at any point in time determines the storage capacity from EVs available to the electricity system. This highlights two of the main differences between stationary battery storage and the use of EVs as battery storage. Firstly, EVs are not available to the system at all times. They are disconnected when driving for instance. However, for the system it is beneficial that the EVs are connected as frequently as possible. Secondly, some of the electricity charged to EV batteries does not go back to the grid but is discharged when the EVs are driven.

The battery storage capacity of EVs, $e_{EV,capacity}$, is in rapid development. According to DNV (Bjørndalen et al., 2023) the average battery capacity of EVs in Norway today is 61 kWh, and they expect that by 2050 this capacity has increased to about 94 kWh. We follow DNV's projections and apply an average battery capacity of 94 kWh for the EVs in the model. The EV batteries' power capacity for discharging ($p_{EV,discharge}$) and charging ($p_{EV,charge}$) is assumed to be identical at 22kW. We assume charging to take place when cars are parked at home or similar locations. Fast charging, charging at high power to recharge the car as quickly and only staying connected as long as charging lasts, is not considered. The reason for this is that such charging behaviour would not facilitate for flexible charging and the flexibility potential of EVs would be lost, at least to some degree depending on the use of fast charging. With only home charging the possible upper limit for charging and discharging power in Norway is 22 kW (Loftås, 2021). However, according to Loftås (2021) most households are today limited to a maximum of 11 kW, but as we are modelling 2050 we assume that all will have the possibility to charge with a power of 22 kW. The efficiency of the EV batteries, $\eta_{charge}$ and $\eta_{discharge}$, is assumed to be 91.49% and 76.91% when charging and discharging respectively. The applied upper charging power capacity of 22 kW assumes a current of 32 A (Loftås, 2021). We apply the EV charging and discharging power losses reported by Apostolaki-Iosifidou et al. (2017) which is measured at 30A. Apostolaki-Iosifidou et al. report a round-trip loss in the EV battery averaged over different SOC at 2.79%. Assuming equal losses for charging and discharging gives a loss of 1.4% from the battery when charging or discharging. The losses from the power electronic unit, which handles the conversion between alternating and direct current, again averaged over various SOC, is reported at 6.79% when charging and 20.21% when discharging. Finally, for the charging station losses are reported at 0.32% when charging and 1.48% when discharging. In total this amounts to losses of 8.51% when charging and 23.09% when discharging, giving efficiencies of 91.49% and 76.91% respectively. Losses are only modelled for charging and discharging, losses caused by storage time is not accounted for as the nature of the EVs makes them most suitable to provide short-term storage, making such losses negligible. The state of charge of the battery is a measure of the energy stored in terms of percentage of its total capacity. Gong et al. (2024) studied how charging strategies, including vehicle to grid, impact the degradation of batteries. They concluded that there is no significant increase in battery degradation from V2G. Nonetheless, it is in general important to charge in a way that minimises the degradation of batteries. In a list of recommendations to ensure long battery lifetime Woody et al. (2020) suggest reducing the amount of time with minimal or maximal state of charge. Thus, to minimise the stress on the batteries and thus also battery degradation the state of charge is constrained to never exceed 80% ($SOC_{max}$) or go lower than 20% ($SOC_{min}$), similarly to Nagel et al. (2024).

When modelling EVs as flexible it is necessary to know how much electricity the EVs use when driving and when it is used. Driving only takes place in the timeframe between when a vehicle disconnects and when it connects to the grid again. Based on this the discharge of electricity from the EV batteries while driving, $e_{EV,driving}(h,z)$, is modelled to be distributed based on the fraction of cars not connected to the grid (Sørensen et al., 2021). The fraction of cars connected to the grid is higher in the weekend than during weekdays, so it follows from this that the discharge from the EV batteries will be higher during weekdays than in the weekend. The more cars disconnected in any given hour, the higher the discharge will be. The calculation of electricity use while driving starts with the total distance driven by passenger cars in each county of Norway in 2023 (Statistics Norway, 2024d). These total distances are then divided by the number of cars in each county the same year (Statistics Norway, 2024a) to get the average driving distance by passenger cars in each county. From the average yearly driving distances the daily average distances are calculated. Daily electricity use while driving per EV is the average daily driving distance per car multiplied with the electricity use when driving, which is assumed at 0.19 kWh/km (Valle, 2021). The hourly average consumption per car while driving for each county, $e_{EV,driving}(h,z)$ is then the daily driving distance per car in each county multiplied with hourly fraction of cars not connected $(1 - c_{EV}(h))$. When EVs are not participating in flexibility they charge in what we call an uncontrolled manner. In terms of the modelling this means that the average charging demand per EV, $e_{EV,static}(h)$, follows static load curves. The load curves are adopted from Sørensen et al. (2022), who provide separate load profiles for workdays, Saturdays and Sundays. From the separate load profiles a weekly normalised load profile is created. Each day's share of the weekly load is calculated from the original load profiles, and this is then used to distribute the weekly total electricity use when driving (calculated using the same approach as previously to ensure that demand from EVs is constant across all scenarios) between the days in the week.

## 2.3  Scenarios

This study considers three different modes of charging for EVs, which vary in the amount of flexibility EVs can provide to the electricity system. Namely, uncontrolled charging, flexible charging and vehicle-to-grid. The three charging modes are detailed in the following:

- **Uncontrolled charging** follows a charging profile corresponding to when EVs typically would be charged. This charging mode provides no flexibility to the system, as the cars are charging according to the profile without consideration of the rest of the system and are thus only seen as a load.
- **Flexible charging (Flex)** allows the model to decide when it is optimal for the system to charge the EVs, thus allowing the model to shift charging loads temporally. This flexibility of when to charge the cars are constrained by a constant driving demand. The EV battery always needs to be charged sufficiently to ensure that the state of charge is high enough to provide the energy needed to fulfil the driving demand.
- **Vehicle-to-grid (V2G)** provides the same temporal flexibility to when the EVs are charged as the flexible charging mode, but in addition it allows the EV batteries to also discharge to the grid. This means that as long as the EVs are connected to the grid, they can be used as battery storage in the electricity system.

In addition to these three charging modes, we model a varying share of the EV fleet providing flexibility to the electricity system. Meaning that they participate in either flexible charging or V2G. The share of participation varies from 0% to 100%, and all vehicles not participating in either of the two flexible charging modes charge according to the uncontrollable charging mode. 0% participation then corresponds to all EVs charging uncontrolled, and 100% entails that every single EV is available for flexibility in the system. The scenarios considered for the model runs are made up of the three charging modes, with uncontrolled charging making up the base scenario, in addition to the varying share of EVs participating in flexible charging or V2G. We model the participation in increments of five percentage points, which results in a total of 21 scenarios: The base scenario and participation in flexible charging and V2G ranging from 5% to 100%. Across all the 21 scenarios the total electricity demand, including the flexible charging of EVs, remains constant. Of these 21 scenarios, three will be given particular attention in the following discussion of the modelling results. These scenarios are of special interest as they represent the extremes of the different modes of charging. With 100% participation in the flexible charging modes, they fully show the effect and potential of the flexibility they provide. These selected scenarios are named as follows:

- *NoFlex*: uncontrolled charging of all EVs, meaning that there is 0% participation in flexible charging and V2G. This is the base scenario, which the flexibility options are benchmarked against.
- *100%Flex*: 100% participation in flexible charging.
- *100%V2G*: 100% participation in V2G.

# 3 Results and discussion

Running the electricity system model for the 21 scenarios on EV charging mode and flexibility participation, as described in section 2.3, provides alternative future electricity system designs for Norway in 2050. The designs illustrate the potential role of EV flexibility in the future electricity system. This section presents the results and discusses the findings from the electricity system optimisations. Starting with the role of EV batteries in the balancing of supply and demand of electricity. Then the impact of EV flexibility on the electricity system's total cost is assessed, before we look at how EV batteries could change the design of the future Norwegian electricity system, with a focus on the participation of EV users.

## 3.1 Supply-demand balance

The main challenge for the system is to generate enough electricity to meet the growing demand, while at the same time ensuring that the supply and demand of electricity is consistently balanced. What we mean with balanced is that in each hour the amount of electricity consumed is exactly equal to electricity supplied, either directly through generation, through transmission, or dispatch from storage. The role which EV batteries could play in the electricity system is best illustrated by looking into how EVs interact with this hourly balancing of supply and demand. For this we consider two cases: The first, as shown in Figure 1, is for a summer day (15$^{th}$ of June) with high generation from solar power. The second case, in Figure 2, considers a winter day (21$^{st}$ of December) with high demand and low supply from solar power. Both figures show the hourly balancing of electricity supply and demand, for the three selected scenarios (i.e. (a) NoFlex, (b) 100%Flex, and (c) 100%V2G). In both cases demand for electricity is divided into three categories: Fixed loads, which remain constant between all scenarios and cannot be shifted temporally. Charging of energy storage, including batteries and pumped hydro, which is optimised for each scenario. EV charging loads, which is fixed in NoFlex but optimised in 100%Flex and 100%V2G. The total demand including EV and energy storage charging is therefore always higher or equal to the fixed demand. Fixed loads are showed as dotted lines, the dashed lines show the charging of storage (which includes stationary batteries and pumped hydro) added to the fixed load, while the solid line also adds the load from EVs.

### 3.1.1 Summer

With the high solar generation in the summer, Figure 1 illustrates how EV flexibility pairs well with variable renewables, in particular solar power. With uncontrollable charging of EVs, it is evident from Figure 1 (a) that charging occurs mainly in the evening from about 16:00 and throughout the night. The uncontrolled charging set in at about the same time as the generation from solar starts to drop off. While the EV charging is not necessarily contributing to increased peak demand, higher loads occur when supply from variable renewables is lower. With the decreasing solar generation and increasing demand from EVs, the variable renewable sources of electricity no longer meet demand. Additional generation from hydropower reservoirs and dispatch from energy storage is needed during these hours. This stationary battery storage is charged during the day, utilising the high PV output, moving this temporally to meet the demand from uncontrollable charging of EVs during the evening and night.

With the aim to integrate high shares of variable renewable energy it is desirable to move the demand to when generation is higher. This would for instance be to the daytime in summer when PV output is high, at least if one wants to avoid deploying additional energy storage such as stationary batteries. This can be achieved with flexible charging and V2G. When allowing the model to decide when it is optimal to charge the EVs, we see in Figure 1 (b) that charging is mostly moved to the daytime in order to utilise the high output from solar PV. In this way one avoids the intermediate step of first charging stationary batteries when solar generation is high and storing it for later in the evening or during the night when it is discharged to meet the charging demand of EVs. However, battery storage is still needed to meet the demand not covered by generation and which cannot be moved temporally.

What happens when EVs are modelled not only to be allowed to charge flexibly but also provide electricity back to the grid, thereby operating similarly to other forms of energy storage such as stationary batteries, is shown in Figure 1 (c). While it looks mostly similar to the 100%Flex scenario in (b), the crucial difference is that the EVs completely replace the stationary batteries. In the scenarios of uncontrolled, NoFlex, and flexible charging, 100%Flex, the stationary batteries are used to move the solar generation temporally from the daytime when solar irradiation is high, to the evening when there is no solar generation. This is also exactly what happens in the 100%V2G scenario, but in this case it is the EVs which are charged in order to send electricity back to the grid in the evening when supply from solar diminish. We also observe for 100%V2G, in (c), that there is no dotted line visible. This is because the dotted and dashed demand lines are exactly equal. There is no charging of energy storage, this is completely replaced by EV batteries. The EVs provide all the battery storage needed by the system.

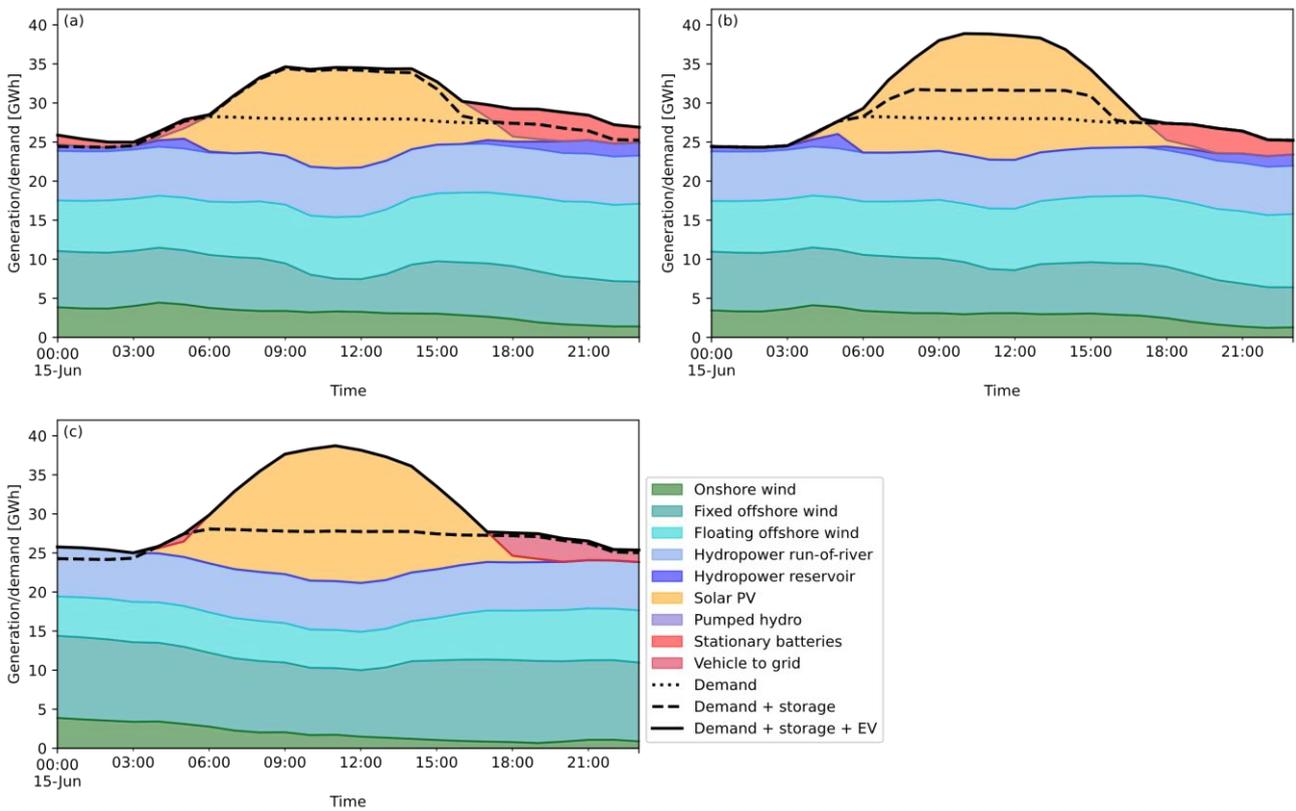

*Figure 1: Balancing supply and demand for a summer day with high solar generation. The aggregated demand and generation of electricity is shown for the three selected scenarios (a) NoFlex, (b) 100%Flex, and (c) 100%V2G. The dotted line shows the fixed loads, the dashed line adds the charging of energy storage, and the solid line shows the total demand with EV charging.*

### 3.1.2 Winter

The supply and demand balance during the winter looks very different to the summer. The main reason for this is the very limited solar resources in Norway during the winter, which is evident in Figure 2, in addition to higher demand. Showing the same scenarios as for the summer, we also here observe how EV flexibility contributes to balancing supply and demand. For the NoFlex scenario, in Figure 2 (a), the higher demand in the winter and lower variable renewable generation this day drives the need for greater dispatch from hydropower reservoirs. We also observe that there is no excess variable renewable generation to charge the two energy storage options, stationary batteries and pumped hydro, as compared to Figure 1 (a). It is however not only EV charging which causes dispatch from energy storage, also other demand throughout the day is met by dispatch.

With flexible charging, as seen in (b), the situation is improved as the EVs are in fact so flexible that charging in full is moved away from this day, which appears to be a particularly difficult one. That not a single EV would be charged for an entire day is not realistic. This is a consequence of the model having the freedom to optimise when to charge the vehicles. The average electric vehicle in the model use about 6.4 kWh of electricity per day for driving. With a battery of 94 kWh a full charge could in theory meet the driving demand for more than fourteen days, illustrating that not charging for one day is in fact very feasible at least for individual EVs. While this reduces the need for energy storage dispatch, especially during the evening and night, the need for stored energy to balance supply and demand is still high. The addition of V2G in (c) again shows how utilising the energy storage capabilities of EVs completely mitigates the need for additional stationary batteries for storage, similarly to what was observed for a summer day. In all three scenarios generation increases towards the end of the day, driven by higher wind power output. In the two first scenarios this goes towards charging stationary energy storage, while with 100%V2G in (c) it is rather used to recharge the EVs.

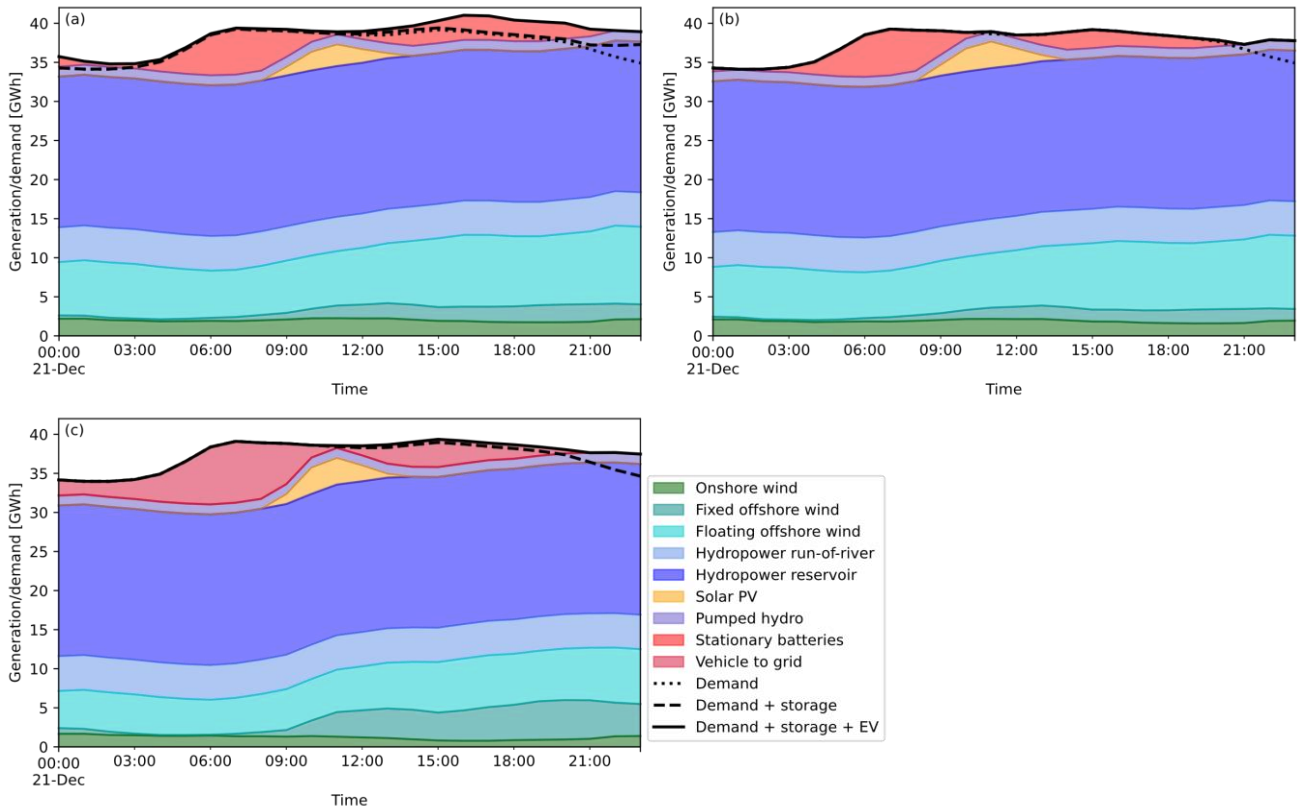

*Figure 2: Balancing supply and demand for a winter day with low solar generation. Three scenarios are considered: (a) NoFlex, (b) 100%Flex, and (c) 100%V2G. Demand is shown with fixed loads, charging of storage and EVs (solid line), only with fixed loads and storage (dashed line), and only fixed loads (dotted line).*

## 3.2  Total system costs

Optimising the system with flexible charging and V2G we observe a declining system cost as the participation rate grows. Figure 3 shows how the objective of the electricity system model, the total system costs, is impacted by EV owners increasingly participating in flexibility, for both flexible charging and V2G. Higher cost savings are obtained with V2G than flexible charging, as expected from the storage capabilities which V2G brings. In the most extreme scenarios, 100%Flex and 100%V2G, the total system cost is reduced by 4.0% and 14.6% respectively, compared to the NoFlex scenario. Especially with V2G there is a major reduction in the system cost. It is evident that V2G provides larger cost savings than flexible charging for all rates of participation. Already at 20% engagement of EV users in V2G the cost benefits are larger than with all EV users engaged in flexible charging. This again shows that using EV batteries for energy storage is much more cost-effective than shifting the charging load. With V2G it is evident that larger cost benefits can be achieved from engaging fewer people. We identified in the introduction that there are many barriers to consumer participation. This makes V2G appealing as it would provide large benefits also at low participation rates, compared to flexible charging, by utilising to a larger extent the potential of the EV batteries.

The total system cost of the electricity system does not decrease linearly as the share of EVs participating in flexibility increase. The addition of flexibility has a larger impact on the system cost when only a small share of the vehicle fleet is flexible. As more and more EVs participate in flexibility the monetary savings obtained with every additional car providing flexibility diminishes. Each additional flexible EV has an incremental smaller impact on the system cost than the last. In other words, the marginal cost savings decrease with each additional flexible EV. This again would mean that each EV user would see less and less value, in terms of cost savings and earnings, from participating the more flexibility it is in the system. This is illustrated by the black dashed lines in Figure 3 which show the slope of the cost curve between 0% and 5% participation for both flexible charging and V2G. We observe the same effect for both modes of charging, even though the effect for V2G is much more prominent. The slopes indicate that at some point the system's need for flexibility could be saturated, having more cars charging flexibly not resulting in additional savings for the system. In such a scenario it would not be cost effective for the system to push for high participation among EV owners. This saturation point is evidently not reached, even at 100% participation additional flexibility would still add value to the system, though with lower cost saving effect.

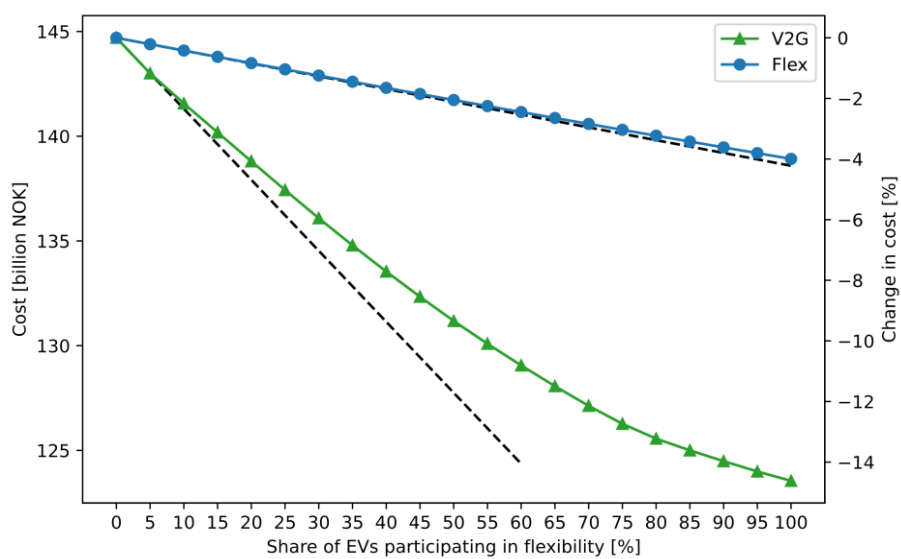

*Figure 3: The cost of the electricity system as a function of the share of EVs participating in flexibility. Both charging modes are considered: flexible charging (Flex) (blue line and circular markers) and V2G (green line and triangular markers). The total system cost in Norwegian kroner is given on the left y-axis, while the right axis shows the percentage change in cost compared to the NoFlex scenario. The dashed lines show the linear extension of the change between 0% and 5% participation.*

In this study the cost of implementing the flexibility from EVs is not taken into account. In that sense the flexibility from EVs is considered free. With this approach we could explore how different rates of participation would impact the system, rather than the system optimising the participation rate based on a set price. The cost of the flexibility could for instance include: The cost of infrastructure to facilitate the participation of EVs in the electricity system (bidirectional chargers for instance), possible payments to EV owners to compensate for losses (such as increased degradation of the EV battery), or the costs of incentives to motivate participation. The results illustrate the savings flexibility from EVs could generate for the system. Whether these savings are enough to cover the costs of the flexibility is up for further investigation. Only from the point of view of costs, there needs to be some savings left after paying the cost of the flexibility. If not, it might not be optimal to build a system depending on consumer flexibility compared to a system providing enough flexibility on the supply side, such as in the NoFlex scenario. However, a clear advantage is the better utilisation of the batteries already available in the EVs instead of investing in additional ones. This point will be discussed further later. From the point of view of the EV owners, participating in flexibility would probably need to bring benefits to the consumer such as cost savings and earnings. With flexible charging this benefit comes in the form of charging the vehicle when the electricity price is the lowest. With V2G, EVs are also charged when the electricity is cheapest, but in addition to that excess energy is sold when prices are highest, thus providing both savings and income. From an economical point of view the flexibility from EVs is only valuable for the system if it decreases the need for other investments in e.g. transmission, energy storage or generation capacity.

## 3.3 Electricity system design

The optimisation shows that the investment in generation and storage capacities is sensitive to the increasing flexibility provided from EVs. Figure 4 shows how the installed power capacities for wind (onshore, fixed offshore and floating offshore), solar and stationary battery storage change with increasing participation in both flexible charging and V2G. Stationary battery storage is by far the most sensitive to the flexibility of EVs. The total capacity of wind and solar varies from 51.2 GW in NoFlex, 53.7 GW in the 100%Flex scenario and 52.2 GW in the 100%V2G scenario. Hydro power capacities are constant between all scenarios. However, changes are more prominent between different model zones and the individual technologies, as the following sub-sections will dive more into detail on.

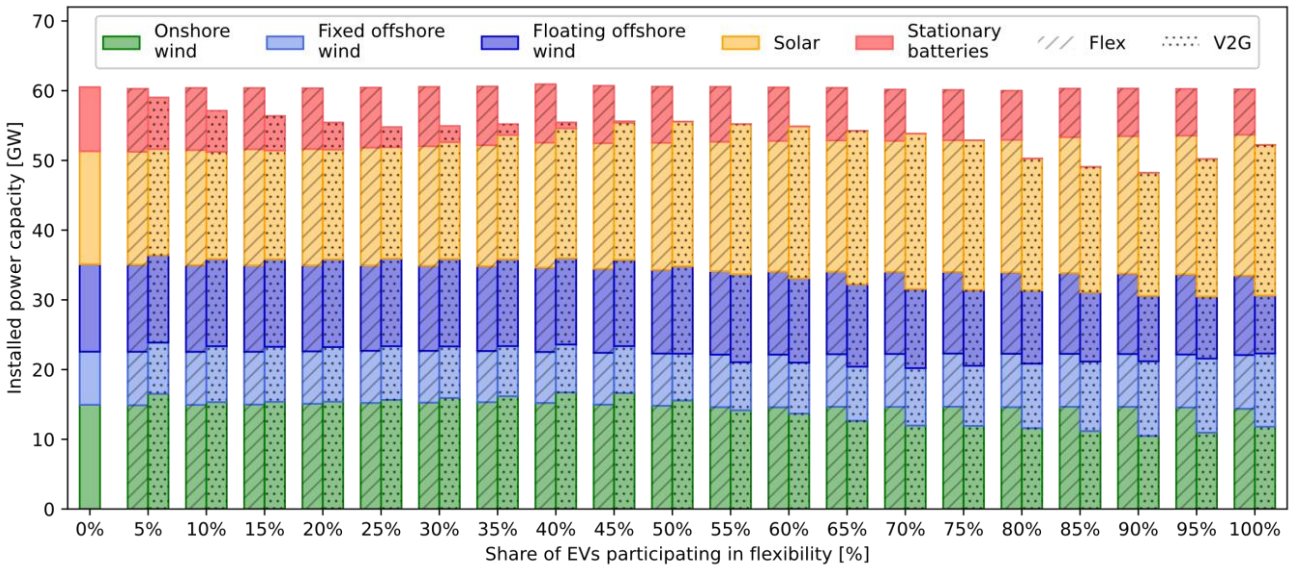

*Figure 4: Installed capacities of variable renewable generation and battery storage power for each scenario. For each participation increment the left bar shows the flexible charging mode and the right bar shows V2G. The bars are stacked.*

Figure 6 shows how the capacities for generation and storage technologies in each zone and the transmission capacities between these zones vary between the three chosen scenarios (i.e. (a) NoFlex, (b) 100%Flex, and (c) 100%V2G). The size of the pie charts illustrates the total installed capacity of generation technologies and battery storage. The larger the pie chart the greater the total capacity. The thickness of the lines illustrates the transmission capacity between model zones, with a thicker line meaning a higher capacity. The zones of the model together with the number of EVs assumed in each zone is given in Figure 5. Table 3 in Appendix C provides the data on the changes in the electricity system design for each zone. Certain technologies have changes of more than 100% for given zones, but as seen in Figure 4 the overall capacity changes in the system are not as substantial.

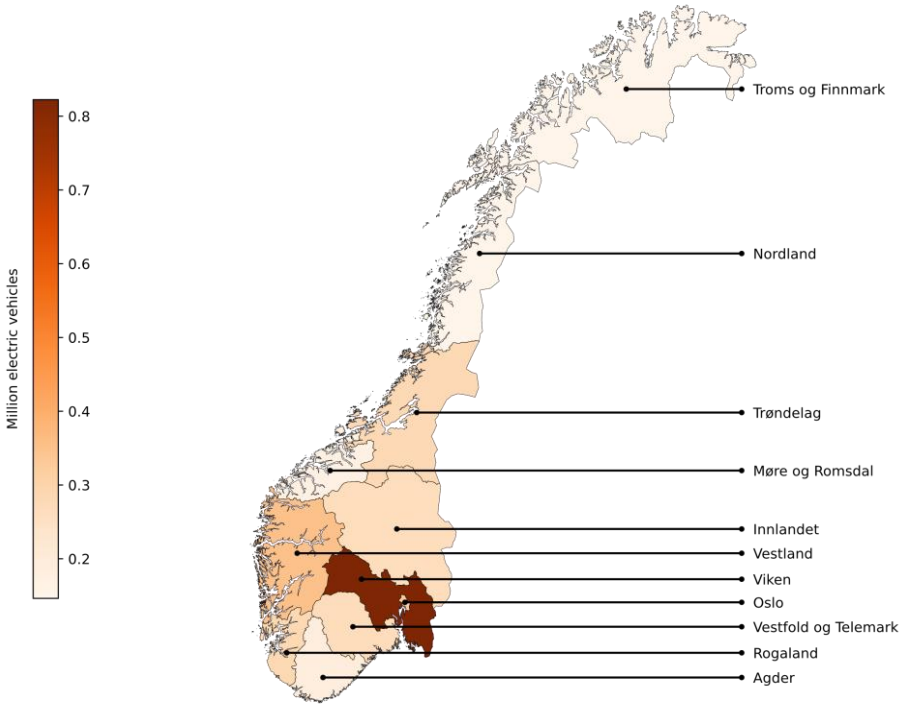

*Figure 5: Number of EVs in each model zone (see Table 2 in Appendix B for details) and the naming of the zones (corresponding to the Norwegian counties). Shapefile from Eurostat (2024).*

The changes between the NoFlex scenario, Figure 6 (a), and the 100%Flex scenario, Figure 6 (b), are evidently quite small. The design in the 100%V2G scenario, Figure 6 (c), is noticeably different from the two other scenarios, with changes in both generation and transmission capacities. The most striking changes in the system design compared to NoFlex occurs in 100%V2G, with the added storage from EVs contributing to integrate high shares of variable renewable generation. There is a slight increase in total variable renewable generation capacity both with 100%Flex and 100%V2G, and generation capacity in general as variable renewable generation technologies are the only investment options in generation for the model. Solar sees an increasing role with the share of solar increasing from 31.6% in NoFlex, to 37.6% and 41.4% respectively in 100%Flex and 100%V2G. This illustrates how EV flexibility helps integrate larger shares of solar generation. In the assumptions of the model solar is the cheapest generation technology the model can invest in, both in terms of investment and operational costs. Battery storage is seen mostly where there is a lot of solar capacity. We observe a shift of solar capacity from Vestfold og Telemark to Viken and Oslo with 100%V2G. With most EV batteries available in Viken, as seen in Figure 5, this shift reduces the need for stationary battery storage. Some counties are largely covered by existing hydropower, this includes Vestland, Innlandet, and Nordland. Due to this they remain largely unchanged between the three scenarios, though Nordland sees some additions of onshore wind. The three northernmost regions of Norway, namely Trøndelag, Nordland and Troms og Finnmark, are the regions seeing the major reductions in transmission capacity. Especially Nordland and Troms og Finnmark becomes less interconnected with the rest of the Norwegian electricity system. This is evident from the difference in the transmission capacity between the two northernmost zones in Figure 6 (a) and (c). For Troms og Finnmark also generation capacity decreases substantially in 100%V2G with no expansion of onshore wind. As both the electricity demand and number of EVs is greatest in southern parts of Norway, this indicates that EV flexibility reduces the need for transmitting electricity from northern to southern parts of Norway.

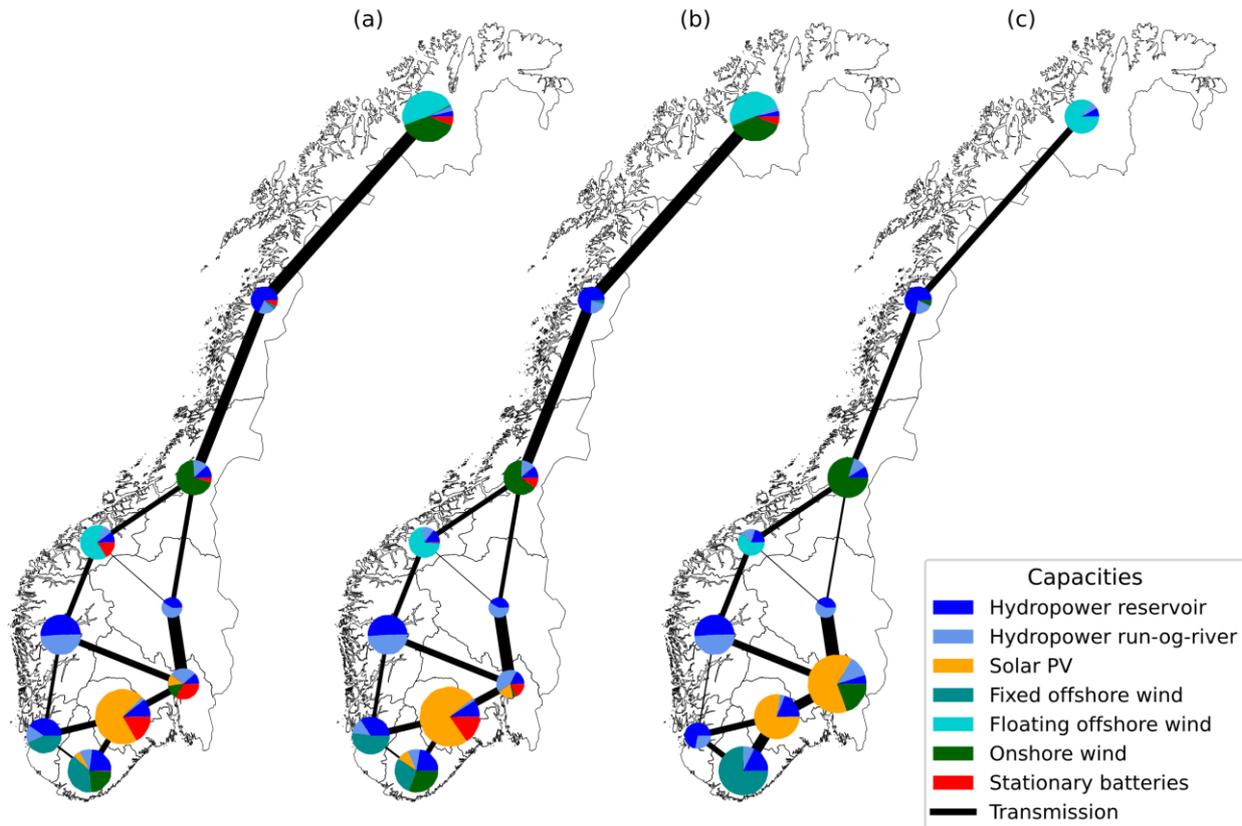

*Figure 6: The electricity system design. The three maps shows how the installed capacity of generation, storage and transmission are distributed spatially between the zones of the model (see Figure 5) for the three selected scenarios: (a) NoFlex, (b) 100%Flex, and (c) 100%V2G. Offshore wind capacity is included in the nearest zone. For clarity Oslo and Viken is merged, as Oslo is situated entirely within and only has transmission capacity with Viken. Only battery storage and solar are built in Oslo. See Table 3 in Appendix C for detailed capacities. Shapefile from Eurostat (2024).*

### 3.3.1 Battery storage

The most significant system design change is observed for stationary battery storage. With increasing participation in EV flexibility, less and less are invested in stationary batteries for energy storage purposes. This is seen in Figure 7 for both flexible charging and V2G. As more and more EVs participate in V2G, more and more battery storage capacity become available to the system. We assume about 3.25 million EVs with an average battery capacity of 94 kWh in Norway in 2050. At 100% participation in V2G this amounts to about 305 GWh of battery storage capacity from EVs made available to the electricity system. We have already seen in Figure 1 for the 100%V2G scenario that for a summer day with high solar PV output there is no longer a need for stationery battery storage. This is however not only the case for that particular day. The results show that with V2G the need for stationary battery storage can be completely eliminated from the system. This occurs already with just 50% of all EV users in Norway adopting V2G, as seen in Figure 7. This illustrates the potential which EV batteries could play in future electricity system with high storage demand.

With V2G the battery storage is not replaced in a relationship of one to one with EV batteries. To replace a certain capacity of battery storage a higher capacity of EV batteries is needed. To replace the 74 GWh of stationary storage in the NoFlex scenario the system needs about a double total EV battery capacity, 152.5 GWh, participating in V2G. This is expected due to the restrictive nature of using EV batteries as electricity storage. The crucial difference between the EV batteries and conventional battery storage is the temporal constraints to when the EVs are available to the electricity system. EVs are not connected to the grid at all times, and parts of the charged electricity must of course go into driving the vehicles. Another factor which limits the degree to which EVs replace stationary batteries is their spatial distribution, as the model cannot optimise where the batteries are located. While more EV batteries than stationary batteries are needed to meet the same storage demand, the EV batteries are already there because they serve the purpose of decarbonising the transport sector. A total of 305 GWh battery capacity is in the system regardless of the flexibility they provide. In the NoFlex scenario the total battery capacity is therefore the 305GWh from the EVs plus the 74 GWh of installed stationary battery storage. In total 379 GWh. With V2G the EV batteries are utilised to a greater degree of their potential. In 100%V2G the total battery capacity is reduced from the 379 GWh in NoFlex to 305 GWh. This is a reduction of 19.5% and it shows there are no stationary battery storage in the system for 100%V2G.

Compared to the V2G scenarios, the stationary battery capacity declines much slower with the increasing participation in flexible charging. In the flexible charging scenarios, the installed stationary battery capacity decrease in a roughly linear manner. The battery capacity is reduced by approximately 1GWh, or 1.5% of the capacity in the NoFlex scenario, for every five percentage points increase in the participation rate. This results in a 29% reduction in the need for stationary battery storage in the 100%Flex scenario compared to NoFlex. The capacity of stationary batteries is also not evenly distributed between the model zones. For the NoFlex scenario stationary battery storage are mainly situated in Vestfold og Telemark where solar capacity is also high. Vestfold og Telemark alone have 38% of the installed capacity, a share which increases to 58% in the 100%Flex scenario.

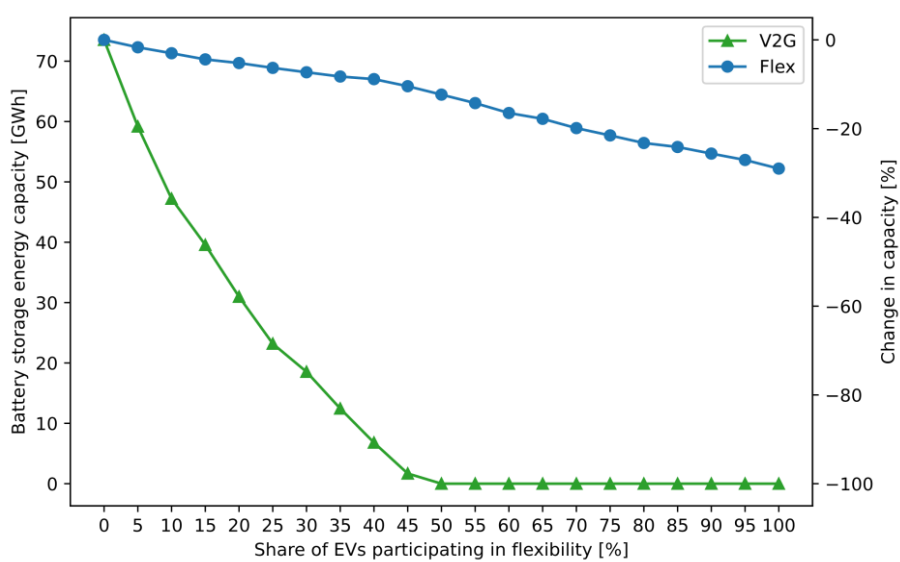

*Figure 7: Stationary battery storage for flexible charging (blue line and circular markers) and V2G (green line and triangular markers) at different participation rates. The figure shows both the total battery storage energy capacity in GWh (left axis) and the relative change compared to NoFlex (right axis).*

EVs are assumed to only charge in the zone which they are registered, any effects of vehicles traveling across model zones is not accounted for. The share of EVs participating in flexibility is also set to be the same across all zones. However, as we see from the system design, Figure 6, for NoFlex the need for electricity storage is greater in certain zones than others. The flexibility from EVs would probably have more effect in these zones. This raises the question of whether aiming for a general uptake of EV flexibility would be best, or if resources could be more efficiently spent targeting areas with higher storage need. Looking into this is however beyond the scope of this paper.

### 3.3.2 Solar power

Solar capacity is increasing with higher participation in both flexible charging and V2G, as seen in Figure 8. Compared to NoFlex the total solar power capacity increases significantly by 25% and 33% for 100%Flex and 100%V2G respectively. This again reflects the way solar power and EV batteries interact as observed when discussing the supply and demand balance in Section 3.1. The model invests in solar as this is the electricity source with the lowest investment, operational and maintenance cost, and because the increasing flexibility from EVs can balance the variability of solar power. The flexible charging scenarios show a relatively steady increase in solar capacity. For the V2G scenarios, on the other hand, we see a drop in capacity compared to NoFlex until about 25% participation. From 25% to 70% participation a steep increase in capacity is seen, reaching a maximum at 70% participation with an increase in capacity of 37.4% compared to NoFlex. At this point however the capacity start dropping off again down to about 8.8% compared to NoFlex at 90% participation. Despite this decrease at 100%V2G the solar capacity sees a steep climb reaching a 33% increase compared to NoFlex. This unexpected behaviour is related to the investment in transmission capacity as seen on the right hand side of Figure 10, and which will be discussed more when considering transmission in Section 3.3.4.

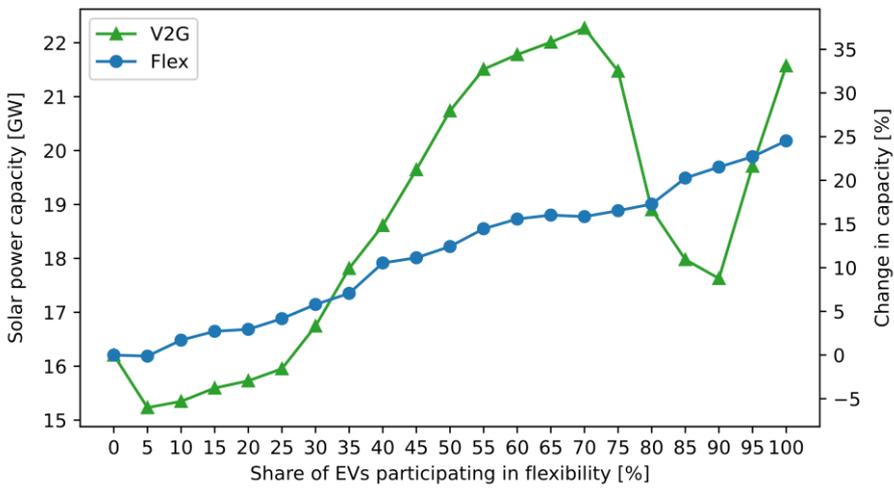

*Figure 8: Solar power capacities for flexible charging (blue line and circular markers) and V2G (green line and triangular markers) at different participation rates. The left axis gives the total installed solar power capacity, while the right axis gives the relative change compared to NoFlex.*

Figure 6 shows how solar fit into the electricity system spatially and illustrate how investments into solar power increase with the added flexibility in 100%Flex and 100%V2G. Solar is mainly installed in the southern parts of Norway. In 100%Flex the major addition of solar capacity is seen in Vestfold og Telemark where there is an additional 25% expansion of solar capacity. Agder sees an additional 67% in solar though capacities were quite small to begin with. Solar is very concentrated in Vestfold og Telemark, in both NoFlex and 100%Flex. 90% of total solar capacity is installed there in both scenarios. While solar capacity is largely concentrated in Vestfold og Telemark in the NoFlex and 100%Flex, it is slightly more distributed with V2G. In 100%V2G solar capacity in Vestfold og Telemark is reduced by about 38% compared to NoFlex, and the capacity is only 50% of that in 100%Flex. This is however compensated with an addition of 11.5 GW of solar capacity in Viken, which has no solar capacity in NoFlex.

### 3.3.3 Wind power

Wind power in total shows a slight decrease in installed capacity when adding flexibility from EVs. Figure 9 shows the capacity as well as the relative change compared to the NoFlex scenario for onshore, fixed offshore and floating offshore wind power, as well as the total wind power capacity, for both flexible charging and V2G. With added flexibility the general trends we observe are an increase in the installed capacity of fixed offshore wind, while floating offshore and onshore wind is decreasing. The total capacity of wind power shows a decreasing capacity with both flexible charging and V2G.

With flexible charging onshore wind, upper left of Figure 9, remains almost unchanged despite the increasing flexibility from EVs. At most, 100%Flex, onshore wind power capacity is less than 3.8% lower than in the base scenario. With V2G onshore wind initially sees some spikes in installed capacity. Noticeably at 5%, 40% and 45% participation capacities are more than 10% higher than in NoFlex. From 45% participation onshore wind capacity starts decreasing, with the lowest point of 29.8% less than NoFlex at 90% participation. For 100%V2G capacity is 21% lower than NoFlex. Fixed offshore wind, upper right of Figure 9, also stays relatively constant with flexible charging, seeing at most a 4.1% decrease compared to NoFlex at 40% participation and a 1.1% higher capacity at 100%Flex. With V2G, fixed offshore wind is the only form of wind seeing a significant increase in any scenario. At most a capacity increase of 40.3% is reached at 90% participation, which reduces slightly to 37.8% for 100%V2G. Floating offshore wind, lower left of Figure 9, sees a steady and relatively linear decrease as more and more EVs charge flexibly. At most the installed capacity is reduced by 9% for 100%Flex. With V2G the capacity remains relatively unchanged compared to NoFlex up to 55% participation. After this floating offshore wind sees a steep decline in installed capacity, with 33.6% less installed capacity for 100%V2G compared to NoFlex. Looking at the wind power capacity in total, lower right in Figure 9, the sum of onshore, fixed and floating offshore, we can see how flexible charging and V2G affect the total investments into wind. Wind power sees a decrease of 4.6% and 12.7% for 100%Flex and 100%V2G respectively. With flexible charging the total wind capacity decreases steadily with increasing participation. With V2G the capacity sees a positive change compared to NoFlex up to 50%. At 50% participation in V2G the total capacity begins to decrease, and more rapidly than for flexible charging. The total capacity decrease is about 1.6 GW for 100%Flex and 4.5 GW for 100%V2G. Compared to solar, which increases capacity with 4.0 GW and 5.4 GW for the corresponding scenarios, we observe that the flexibility from EVs enable the transition from wind to solar power. This transition is driven by solar being less expensive than wind power and the model being cost optimising.

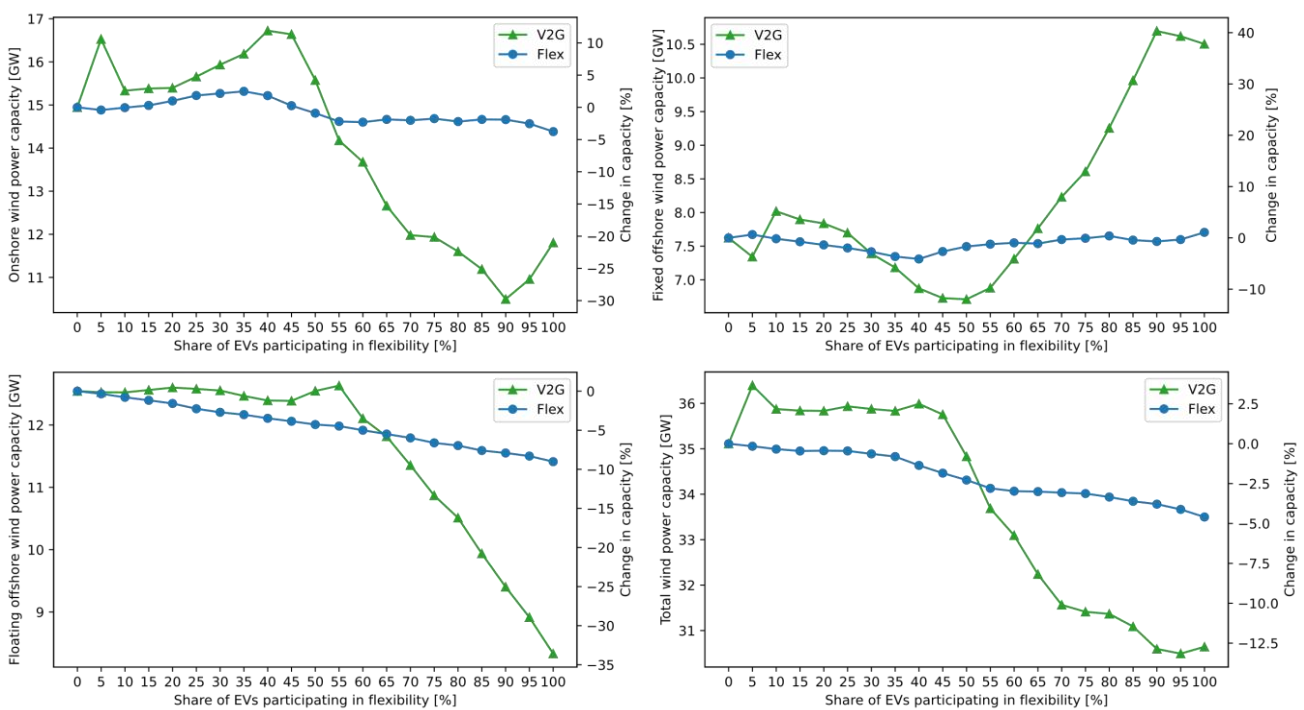

*Figure 9: Power capacities for onshore wind (top left), fixed offshore wind (top right) and floating offshore wind (bottom left). The total wind power capacity (the sum of onshore, fixed offshore and floating wind) is shown in the final figure (bottom right). Results are shown for flexible charging (blue lines and circular markers) and V2G (green lines and triangular markers) for all participation scenarios. The left axes show the installed capacity in GW and the right axes show the relative change in capacity compared to NoFlex.*

In Figure 6 we see the spatial distribution of wind power for the three selected scenarios. For 100%Flex onshore wind sees a minor decrease in capacity with the added flexibility. Onshore wind is primarily found in Troms og Finnmark and Trøndelag for NoFlex, which remains the case also for all participation rates of flexible charging. While the total changes are small, some zones see major additions and reductions in capacity. Noticeably, 83% less onshore wind in Viken while there is 36% more in Agder. Troms og Finnmark sees a decrease of approximately 7% but remains the zone with highest onshore wind capacity. The total capacity of fixed offshore wind is largely unchanged between NoFlex and 100%Flex. The main observed change is a shift of capacity from Agder to Rogaland which see a 47% increase in fixed offshore wind capacity. In all three of the selected scenarios floating offshore wind is only built in Møre og Romsdal and Troms og Finnmark. For 100%Flex the total floating offshore wind capacity decreases by about 9%. This decrease mainly happens in Møre og Romsdal where installed capacity decreases by 23% compared to NoFlex.

For 100%V2G both onshore and floating offshore wind become less important in the system, while fixed offshore wind capacity increases. For onshore wind total capacity decreases by almost 21%. Two zones which had noticeable capacity for NoFlex see a 100% reduction with 100%V2G, namely Agder and Troms og Finnmark. In Viken onshore capacity is more than tripled compared to NoFlex, and in Trøndelag onshore wind increased by 49%. For fixed offshore wind, the capacities become a lot less spatially distributed. Only one zone has installed capacity, and three zones lose all capacity compared to NoFlex. Agder, now accounting for the whole capacity, see fixed offshore wind increase by 156%. Floating offshore wind still is only present in Møre og Romsdal and Troms og Finnmark in 100%V2G. However, capacity decreases by 52% and 20.5% in the two zones respectively. The total capacity of floating offshore wind is reduced by almost 34% between NoFlex and 100%V2G.

### 3.3.4 Transmission

Transmission is a source of flexibility balancing the spatial variability of renewables. The left graph in Figure 10 shows the total installed transmission capacity for all scenarios. With flexible charging we observe that higher participation in EV flexibility results in higher transmission capacity. This is however a very minor increase of no more than 2.4% for 100%Flex compared to NoFlex. Interestingly with V2G the opposite trend is observed, with transmission capacity decreasing as more flexibility is added. With V2G we first observe an increase of up to 2.3% compared to NoFlex at 25% participation, before the total transmission capacity falls abruptly reaching the lowest point of 13.8% less than NoFlex at 65% participation. At 90% participation it increases again up to only 4.8% less than NoFlex, before falling to 7.7% less for 100%V2G. This decrease in transmission capacity seems reasonable due to the energy storage possibility obtained with V2G. It might not be expected to see this opposite behaviour between flexible charging and V2G as the main difference between the two is that V2G just adds additional flexibility. However, this is all connected to the rest of the decisions the model makes with the generation and storage capacities. With lower transmission capacity supply and demand must be balanced within the zones to a greater extent. The added flexibility from EVs, which are distributed to each model zone as seen in Figure 5, supports this balancing. Every zone has larger opportunities to balance supply and demand without imports and exports through the transmission grid. More flexibility within each zone means that the ability to integrate more of the locally generated electricity is higher, reducing the need to invest in transmission.

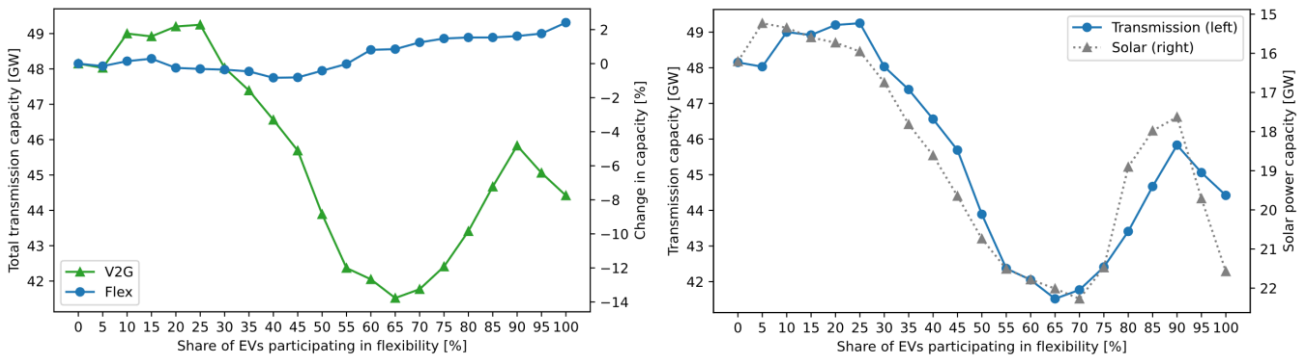

*Figure 10: The left-hand part shows the total transmission capacity in GW (left-hand axis) and the relative change compared to NoFlex (right-hand axis) for all participation rates in flexible charging (blue line and circular markers) and V2G (green line and triangular markers). The right-hand part shows the total transmission capacity for all participation rates in V2G (left-hand axis) compared to the inverted total solar power capacity (right-hand axis).*

The change in total transmission capacity between the different participation scenarios shows an unexpected development with V2G. The change in capacity is not steady as with flexible charging, but several local maximums and minimums are observed. As mentioned earlier it is observed that there is a close relation between the total capacity of solar power and transmission. Therefore, on the right-hand side of Figure 10 we have plotted the transmission capacity together with the inverted solar capacity for all participation scenarios of V2G. From this it becomes clearer that the valleys in the transmission capacity corresponds very well to the peaks in solar capacity and vice versa. With more solar capacity the model invests less in transmission capacity and vice versa. This might be counterintuitive as one would assume more variable generation, such as solar, would require more flexibility, from transmission for instance. The case is however that solar is built increasingly in the zones with most EVs, and thus greatest flexibility capacity.

The spatial distribution of the transmission capacity is seen in Figure 6. For 100%Flex transmission is largely unchanged compared to NoFlex, increasing in total only 2.5%. Only marginal changes appear between the different zones. This corresponds well with the relatively minor changes in generation and storage capacities for 100%Flex as well. Larger changes in the electricity system design are seen for 100%V2G. Total transmission capacity is reduced by almost 8% compared to NoFlex, but the changes in transmission capacity between certain zones are larger. In Vestfold og Telemark and Agder transmission capacities are increased by 45% and 89% compared to NoFlex. Transmission capacity in Trøndelag, Nordland, and Troms og Finnmark is reduced by 37%, 46%, and 49% respectively. The reduced transmission capacity is linked with the decreased generation capacities in the two latter zones, as less electricity is transmitted southwards. We allow for transmission expansion which, as another flexibility option, compete with EVs providing flexibility. Thus, by limiting transmission expansion the effects of EV flexibility would likely be even stronger. However, in our approach transmission expansion has an associated cost, while the EV flexibility is seen as free by the model.

# 4  Conclusion

The electrification of passenger cars creates opportunities for operating EVs as flexible loads by shifting charging to times which benefits the system. Further, the batteries in EVs can be utilised as energy storage when the EVs are not driven. Both these options enable EVs to contribute to the integration of variable renewable energy, essential to reaching climate targets. Due to uncertainty in the participation in flexibility by EV users, this work has optimised future net-zero Norwegian electricity systems applying various charging modes and participation rates in these. We provide an in-depth answer to the research question: How is the design and cost of a future Norwegian net-zero electricity system affected by the share of electric vehicle users participating in flexible charging or vehicle to grid?

In our analysis we have shown how demand-side flexibility from EVs contributes significantly to reducing the cost of designing a 2050 Norwegian electricity system. This in spite of the fact that the Norwegian electricity system has large resources of energy storage and dispatchable generation through hydro power. Increased demand due to electrification makes additional flexibility essential to integrate variable renewable energy. The two flexible charging modes, flexible charging and vehicle-to-grid, both provided benefits to the system, but the latter much more than the former. Vehicle-to-grid offers greater cost savings for the electricity system compared to flexible charging across all levels of participation. Even with just 20% engagement of EV users in V2G, the system's cost benefits surpass those achieved by engaging all EV users in flexible charging. This emphasises the significant cost-effectiveness of utilising EV batteries as energy storage. It demonstrates that substantial savings can be realized with a relatively small subset of participants, rather than requiring broad engagement in load-shifting strategies. The optimisation showed that vehicle-to-grid made the investment in additional stationary battery storage redundant already at 50% participation. Half of the EV fleet could provide all the additional storage the system needs. The additional energy storage available to the system at 100% participation in vehicle-to-grid also enables expensive, and controversial, wind power expansion to be substituted by cheaper solar power. Investments in transmission to provide spatial flexibility to the system is also reduced compared to the case with no flexibility from EVs.

Our results show that costs savings from EV flexibility are substantial, which on the other hand, could be used to incentivise and compensate users to provide flexibility and install the necessary infrastructure. While the results show that engaging 100% of EV users lead to the cheapest system design, they also suggest that it may not always be optimal to pursue maximal engagement, particularly when certain consumers could be hard-to-reach. The cost savings increase for every additional vehicle participating, but the savings per vehicle decrease as more participate in flexibility. In such cases, the costs associated with incentivising or compensating these hard-to-reach consumers may exceed the benefits gained from their participation. Also, as higher system benefits are achieved at lower participation rates with vehicle-to-grid compared to flexible charging, the incentives might be better spent on already engaged consumers to motivate participation in vehicle-to-grid, rather than on hard-to-reach consumers to engage them in flexible charging. The analysis here focused on a modelling study of Norway, we suspect that the conclusions on the role of EVs and hard-to-reach consumers would apply to other regions of the world as well. Many regions don't have the benefit of flexible hydro power as in Norway and might show stronger benefits of EV flexibility as transport sectors and energy systems are electrified.

## 4.1 Limitations and suggestions for future work

The substantial impact which participation have on the potential of EV flexibility warrants further research on the topic. In order to design electricity systems which fully benefits from EV flexibility it is essential with a thorough understanding of the amount of flexibility to expect. Our findings show that the electricity design varies significantly depending on the flexibility provided. Several factors impact this but one we would like to point out specifically is the future distribution of EVs. We based the number of EVs on a prognosis expecting growth in the number of vehicles and assumed the distribution to be as currently. If more sustainable modes of transport are adopted, for instance a shift to more public transport, this would alter the amount of flexibility EVs could provide. Further research into how future mobility behaviour could affect the EV flexibility potential and the electricity system would be beneficial. However, our results already provide an indication that lower levels of participation still benefit the system significantly. Another major uncertainty is the charging behaviour which determines to what degree EVs are available in the grid. We based the model on a case study on home charging with long connection times. How EVs charge in the future is highly uncertain, and a shift towards fast charging would limit the flexibility potential. With fast charging EV connection times could be reduced to only the time needed for charging. Longer connection times on the other hand, for instance when charging over night at home, enables much greater flexibility potential.

We model the flexibility from EVs as free as we focus on the rate of participation. However, including the cost, which could come from the need for infrastructure (e.g. bidirectional charging stations) and incentives to motivate participation, could be used to determine the optimal participation levels. Several studies also show that there is a high potential for other sources of demand side flexibility in Norway (e.g. studies by Söder et al. (2018), and Kirkerud et al. (2021)), and EVs might face competition in providing flexibility. Norway could end up in a situation with more flexibility than needed. Already with a lot of flexibility from EVs we observe signs of saturation where the cost savings of additional flexibility is diminishing. Ultimately this could mean that Norway could contribute to supporting other countries with less flexibility potential to integrate large shares of variable renewable energy. Other sources of flexibility or the interconnection of Norway to the European electricity system was not in the scope of this study. It could be an interesting angle for future research, putting the different sources of flexibility up against each other and investigating the role of Norwegian flexibility in the larger European system.

# Acknowledgements

We acknowledge funding from EMPOWER, funded by UiO:Energy and Environment and dScience, and NUANCE, funded by Nordic Energy Research through the Norwegian Research Council.

We would like to acknowledge Tatjana Krackhardt's review of the literature on hard-to-reach consumers as part of a summer project funded by UiO:Energy and Environment.

The development of the EV module started as part of a research project in the course TEK9410 – Energy Markets and Regulation – Electricity System Modelling and Analysis at University of Oslo. This EV module was further developed, including the addition of the share of participants, and implemented in highRES as part of the present study.

## Author contributions

**Tobias Verheugen Hvidsten:** Conceptualization, data curation, formal analysis, investigation, methodology, software, validation, visualization, writing – original draft, writing – review and editing. **Maximilian Roithner:** Conceptualization, data curation, investigation (PV modelling), methodology, software, writing – review and editing. **Fred Espen Benth:** Supervision, writing – review and editing. **Marianne Zeyringer:** Conceptualization, funding acquisition, supervision, writing – review and editing.

## Code and data availability

All code and data are available upon request from the corresponding author.

## Appendix A: Electricity demand

The Norwegian electricity demand in 2050 is assumed to be 260 TWh based on the high electricity consumption prognosis from Statnett (2023). Here we describe the method to distribute this demand spatially between the model zones, shown in Figure 5, and to an hourly temporal resolution. The method we apply to model the demand has been partially described previously in a report from Hansson et al. (2023) and in the paper by Roithner et al. (2024). The following section supplement these previous descriptions.

Interannual electricity demand from Frysztacki et al. (2022) is applied as the base demand onto which future demand from the various sectors are added. A decrease of 7 TWh in general consumption and losses, which is mostly heating (Statnett, 2023), is applied as a constant percentage loss across all hours. Electricity demand from electric transport increase by 23 TWh. With the EV module described in section 2.2 battery electric passenger cars add about 8.4 TWh to the total demand. The remaining 14.6 TWh are distributed spatially as previously described in the report from Hansson et al. (2023). Electricity demand from industry adds 70 TWh, and is spatially distributed based on electricity use by industry in 2019 (Statistics Norway, 2024c). The petroleum sector adds 6 TWh. However, an overview of the plans for electrification of the petroleum sector from Hovland (2023) shows yearly electricity demand amounting to 16.85 TWh. The 6 TWh is therefore distributed based on each zone's share of the demand from the electrification projects.

Demand from battery production and data centres adds 29 TWh (Statnett, 2023). The same four battery factories as previously described by Roithner et al. (2024) is assumed, but with slightly different production capacities and electricity demand. For the battery factory in Rogaland (Beyonder, 2022) we assume a production capacity of 40 GWh, comparable to the other factories. For the remaining three we have 29 GWh in Nordland (FREYR Battery, 2022), 43 GWh in Agder (Morrow Batteries, 2023), and 40 GWh in Trøndelag (Elinor Batteries, 2023). We assume 65 GWh of electricity demand per GWh battery production capacity based on Kurland (2019). For data centres there are planned capacities of 240 MW in Rogaland (Røise, 2023), 150 MW in Innlandet (Vogt et al., 2023), and 840 MW in Vestfold of Telemark (Torstveit, 2024). Data centres could use up to 80% of their maximum capacity (Torstveit, 2024) and we assume them to operate at this level through the year. These battery production and data centre projects amount to about 18.5 TWh of the total 29 TWh. The remaining 10.5 TWh is distributed evenly between the zones. The loads from industry, data centres and battery factories are assumed to be continuous, so the demand is distributed equally across all hours. This is in line with how other battery factories are operated (Kurland, 2019).

# Appendix B: Electric vehicle distribution

*Table 2: Distribution of cars between the model zones. The projected number of EVs in Norway in 2050 (Fridstrøm, 2019) distributed based on the number of cars in each county of Norway in 2023 (Statistics Norway, 2024a).*

| Zone (county name) | Number of EVs in 2050 |
|---|---|
| Oslo | 324 154 |
| Rogaland | 283 611 |
| Møre og Romsdal | 172 388 |
| Nordland | 151 072 |
| Viken | 822 426 |
| Innlandet | 263 113 |
| Vestfold og Telemark | 266 854 |
| Agder | 187 489 |
| Vestland | 351 529 |
| Trøndelag | 278 712 |
| Troms og Finnmark | 146 291 |
| Total | 3 247 639 |

# Appendix C: Detailed capacities

*Table 3: Installed capacities of battery storage, solar, onshore wind, fixed offshore wind, floating offshore wind, and transmission, for each model zone for three selected scenarios: (a) NoFlex, (b) 100%Flex, and (c) 100%V2G. All values are in GW except for the battery storage which is given in GWh.*

|  | Battery storage | | | Solar | | | Onshore | | | Fixed offshore | | | Floating offshore | | | Transmission | | |
|---|---|---|---|---|---|---|---|---|---|---|---|---|---|---|---|---|---|---|
|  | (a) | (b) | (c) | (a) | (b) | (c) | (a) | (b) | (c) | (a) | (b) | (c) | (a) | (b) | (c) | (a) | (b) | (c) |
| Oslo | 20.7 | 8.3 | 0 | 1.0 | 1.0 | 1.0 | 0 | 0 | 0 | 0 | 0 | 0 | 0 | 0 | 0 | 3.0 | 3.0 | 3.0 |
| Rogaland | 0 | 0 | 0 | 0 | 0 | 0 | 0 | 0 | 0 | 3.0 | 4.4 | 0 | 0 | 0 | 0 | 4.2 | 4.9 | 4.6 |
| Møre og Romsdal | 10.8 | 0.5 | 0 | 0 | 0 | 0 | 0 | 0 | 0 | 0 | 0 | 0 | 5.2 | 4.0 | 2.5 | 6.5 | 6.5 | 6.5 |
| Nordland | 3.0 | 0 | 0 | 0 | 0 | 0 | 0 | 0 | 0.3 | 0.2 | 0.2 | 0 | 0 | 0 | 0 | 13.4 | 13.8 | 7.2 |
| Viken | 1.3 | 0 | 0 | 0 | 0 | 11.5 | 1.2 | 0.2 | 3.9 | 0 | 0 | 0 | 0 | 0 | 0 | 17.9 | 18.4 | 19.2 |
| Innlandet | 0.1 | 0 | 0 | 0 | 0 | 0 | 0 | 0 | 0 | 0 | 0 | 0 | 0 | 0 | 0 | 10.3 | 10.0 | 8.7 |
| Vestfold og Telemark | 28.0 | 30.5 | 0 | 14.6 | 18.2 | 9.1 | 0 | 0 | 0 | 0 | 0 | 0 | 0 | 0 | 0 | 8.2 | 9.0 | 11.9 |
| Agder | 0.4 | 0 | 0 | 0.6 | 1.0 | 0 | 2.5 | 3.4 | 0 | 4.1 | 3.1 | 10.5 | 0 | 0 | 0 | 4.6 | 4.3 | 8.7 |
| Vestland | 0 | 0 | 0 | 0 | 0 | 0 | 0 | 0 | 0 | 0 | 0 | 0 | 0 | 0 | 0 | 9.0 | 9.3 | 7.7 |
| Trøndelag | 2.6 | 6.3 | 0 | 0 | 0 | 0 | 5.1 | 5.1 | 7.6 | 0 | 0 | 0 | 0 | 0 | 0 | 12.0 | 12.1 | 7.6 |
| Troms og Finnmark | 6.7 | 6.6 | 0 | 0 | 0 | 0 | 6.1 | 5.7 | 0 | 0.3 | 0 | 0 | 7.3 | 7.4 | 5.8 | 7.2 | 7.3 | 3.7 |
| Total | 73.5 | 52.2 | 0 | 16.2 | 20.2 | 21.6 | 14.9 | 14.4 | 11.8 | 7.6 | 7.7 | 10.5 | 12.5 | 11.4 | 8.3 | 48.1 | 49.3 | 44.4 |

# Appendix D: Solar PV modelling

For this study the modelling of PV generation was improved by using satellite data from the recently released SARAH-3 dataset (Pfeifroth et al., 2023). Babar et al. (2019) compare satellite and reanalysis datasets for solar radiation. They find that satellite data like SARAH provide the most accurate data. However, the SARAH dataset is only available up to a latitude of 65°N, but Babar et al. (2019) suggests that ERA5 can be applied to supplement the missing data in SARAH. Since SARAH-3 only cover Norway up to roughly the northern border of Trøndelag we apply satellite data from SARAH-3 where available and resort to reanalysis data from ERA5 (Hersbach et al., 2023) for the northern part of Norway.